# Benchmarking Waitlist Mortality Prediction in Heart Transplantation Through Time-to-Event Modeling using New Longitudinal UNOS Dataset

Yingtao Luo, MS[1], Reza Skandari, PhD[2], Carlos Martinez, MS[3],
Arman Kilic, MD[4], Rema Padman, PhD[1]
[1]Carnegie Mellon University, Pittsburgh, PA, USA; [2]Imperial College, London, UK; [3]United Network for Organ Sharing, USA; [4]Medical University of South Carolina, Charleston, SC, USA

**Abstract**

*Decisions about managing patients on the heart transplant waitlist are currently made by committees of doctors who consider multiple factors, but the process remains largely ad-hoc. With the growing volume of longitudinal patient, donor, and organ data collected by the United Network for Organ Sharing (UNOS) since 2018, there is increasing interest in analytical approaches to support clinical decision-making at the time of organ availability. In this study, we benchmark machine learning models that leverage longitudinal waitlist history data for time-dependent, time-to-event modeling of waitlist mortality. We train on 23,807 patient records with 77 variables and evaluate both survival prediction and discrimination at a 1-year horizon. Our best model achieves a C-Index of 0.94 and AUROC of 0.89, significantly outperforming previous models. Key predictors align with known risk factors while also revealing novel associations. Our findings can support urgency assessment and policy refinement in heart transplant decision making.*

## Introduction

Heart failure is a major global health challenge, imposing a significant burden on patients, clinicians, and healthcare systems (1). In the United States, heart disease has been the leading cause of death since 1950, with approximately 162.1 deaths per 100,000 population in 2023 (2). Globally, ischemic heart disease also remains the top cause of death since 2011, accounting for around 8 million deaths annually (3). About 65 million adults worldwide are living with heart failure, a number expected to rise with aging populations (4). Despite advances in treatment and transplantation, outcomes remain poor, particularly for end-stage heart failure patients, where mortality can reach 75% within one year without interventions such as device implantation or heart transplantation (5).

In the United States, heart transplantation is overseen by the United Network for Organ Sharing (UNOS) and the Organ Procurement and Transplantation Network (OPTN). Historically, the U.S. heart allocation system relied on broad priority tiers (e.g., UNOS Status 1A/1B/2), which were updated infrequently and failed to capture rapid changes in patient condition (6). In 2018, UNOS implemented a revised six-tier allocation system aimed at improving risk stratification and reducing waitlist mortality. Although this revision marked progress, recent studies show that the current framework still only moderately distinguishes patients at highest risk of death before transplant (7–9). Some status categories even display overlapping or inconsistent mortality risk levels, emphasizing the limitations of relying on static, categorical designations (10).

Existing waitlist mortality prediction models in the literature (11–19) have been limited in two key ways. First, they rely heavily on patient variables recorded only at the time of listing, which does not reflect how a patient's condition may change over time. Key prognostic indicators, such as estimated glomerular filtration rate (eGFR) and serum albumin levels, can fluctuate over time or deteriorate due to complications like device malfunctions, making one-time assessments insufficient. This underscores the need for time-sensitive, dynamic models that continuously update patient risk based on evolving longitudinal clinical information (20). Second, despite decades of transplant data and modeling efforts, there is no established benchmark for evaluating the accuracy of waitlist mortality prediction models in the U.S. context. Prior studies rarely report standard performance metrics such as the concordance index (C-index) or AUC (Area Under the Receiver Operating Characteristic Curve), making it difficult to assess whether the existing models are satisfactory. One exception—the French Candidate Risk Score—showed a concordance index (C-Index) of 0.71 (21), which is still inadequate for high-stakes decision-making. While single-center studies on a small cohort of 414 patients have demonstrated the feasibility of continuous risk estimation (22), substantial state-level variations in waitlist mortality exist, and large-scale, nationally benchmarked models are needed but remain lacking.

Encouragingly, the 2018 UNOS policy revision also introduced more granular and frequent data collection, capturing sequential updates in patient status (e.g., device changes, laboratory values, and clinical events). This rich longitudinal data enables the development of more accurate predictive models that treat patient risk as a moving target. This new structure offers a valuable opportunity to apply modern machine learning techniques to build time-to-event models that dynamically update survival predictions in response to evolving clinical conditions. In this study, we leverage these longitudinal data streams to build a time-to-event machine learning model that updates each patient's survival probability in response to new clinical information. Unlike prior models, our approach can incorporate longitudinal changes, such as newly recorded hemodynamic measurements, which may drastically affect a patient's waitlist risk.

We aim to address two critical gaps simultaneously: the lack of benchmarks and the absence of dynamic models. We accomplish this by benchmarking multiple waitlist mortality prediction models, including linear, machine learning, and deep learning approaches, using new sequential waitlist data collected by UNOS since 2018. Our models leverage 77 clinical variables in total from 23,807 patients to estimate survival both at the population level and for individual trajectories. We evaluate performance using survival curves (e.g., concordance index) and 1-year discrimination (e.g., time-dependent AUC), demonstrating substantially improved predictive performance against previous models trained on variables recorded only at listing. In addition, we identify key predictors of waitlist mortality, including established clinical risk factors and novel associations revealed by longitudinal data, and validate these findings against existing literature. By harnessing this comprehensive dataset with modern machine learning techniques, we aim to benchmark the waitlist mortality prediction models, deliver a more accurate data-driven tool that supports clinicians in assessing patient urgency, informs transplant allocation decisions, and ultimately reduces preventable waitlist deaths.

**Methods**

*Data Statistics and Preprocessing:* The data was obtained from the UNOS database, which collects all data related to waiting list registrations and transplantation that have been listed or performed in the United States and reported to the Organ Procurement and Transplantation Network (OPTN) since October 1, 1987. The UNOS database is a national-level organ allocation data source. The subset of interest in the UNOS database is the thoracic dataset (for heart transplantation), which contains a rich set of longitudinal variables of patients who have been waitlisted since October 18, 2018, and the complete donor-patient match history. For this paper, we queried the latest UNOS database that updates its last record on December 31, 2023, for all the patients that were enlisted. Of the 129,616 patient records, only 23,807 with dynamic waitlist history were included in the analysis, as the remainder contained only information from the time of initial listing. Of the 23,807 records included in this analysis, 17,358 patients (72.9%) received a transplant while on the waitlist, 1,898 (8.0%) died while waiting, and 4,551 (19.1%) were removed from the waitlist. For the longitudinal data, the average time to update patient waitlist record is 720 days (SD 2570), with a median of 92 days. We used Kaplan-Meier survival curves to estimate population-level survival probabilities over time. This non-parametric method allows visualization of survival probabilities.

Time zero for this study was the initial date of active UNOS listing for heart transplantation. The endpoint was all-cause mortality before heart transplantation, with follow-up until December 31, 2023. Patients were censored at the time of transplantation. Patients alive at the end of the study without receiving a heart transplant should have been censored at that time; however, no such patients were identified in the cohort with dynamic waitlist history. Patients lost to follow-up after time zero were censored at the last documented encounter within the UNOS database system. Patients with multiple waitlist records due to transfers between transplant centers had their records consolidated into a single comprehensive record. For patients with multiple waitlist records due to reasons other than center transfers (such as clinical improvement followed by subsequent re-listing), only the most recent waitlist record was included. Delisting was not considered a censoring event, in accordance with the intent-to-treat principle; thus, patients who experienced clinical deterioration to the point of transplant ineligibility were neither excluded nor censored from the analysis. To handle missing values, baseline variables with missing values of 30% or greater were not included. Variables with less than 30% missing values had missing data imputed with 5-fold multiple imputation using a Markov Chain Monte Carlo technique to obtain final parameter estimates and the imputed data. To address multicollinearity, we calculated the Variance Inflation Factor (VIF) and excluded variables with VIF > 10 to implement Cox model. Categorical variables were one-hot encoded, and numerical variables were standardized using z-score normalization.

*Waitlist Mortality Time-to-Event Modeling Benchmark:* In the literature, several papers have reported modeling mortality risk among patients listed for heart transplantation using the data recorded at the time of listing. Alshawabkeh et al. (11) used Fine and Gray competing risks models to estimate death or delisting due to worsening in 1,290 adults with congenital heart disease (ACHD) and 38,557 non-ACHD adults from the Scientific Registry of Transplant Recipients (SRTR) dataset. Jasseron et al. (21) proposed a simplified risk score model derived from Cox regression, using a few baseline variables (e.g., eGFR, bilirubin, BNP) for estimating mortality risk in the French registry data with 2,615 patients. Hsich et al. (13) used Random Survival Forests (RSF), a nonparametric ensemble method, to model waitlist mortality and identify complex interactions between predictors such as renal function, albumin, and mechanical support. The model captured nonlinear interactions across a large national cohort of 49,025 adults. Most recently in 2020, Bakhtiyar et al. (12) employed standard Cox proportional hazards models stratified by ventricular assist device (VAD) use and allocation time eras to evaluate historical mortality trends using UNOS registry data of over 95,000 candidates. Additionally, one notable study by Blackstone et al. (13) employed a time-dependent Cox proportional hazards model using single-center longitudinal data from 414 heart transplant candidates at the Cleveland Clinic. However, due to the use of center-specific variables not available in the national UNOS dataset, their results are not directly reproducible in our setting.

*Time-Dependent Cox Proportional Hazard Model:* We employed a time-dependent Cox proportional hazards model (23) as our own baseline model to analyze waitlist outcomes. Mathematically, such a model is expressed as:

$$h\left(t \middle| x(s)\right) = h_0(t)\exp((x(t) - \bar{x})'\beta)$$

where $h\left(t \middle| x(s)\right)$ represents the cause-specific hazard at the time $t$ given the time-dependent covariates $x(s)$, $h_0(t)$ is the baseline hazard function, $\bar{x}$ represents the mean values of covariates, $s$ denotes the time when the events happen and covariates $x$ update, and $\beta$ is the vector of regression coefficients. This time-dependent framework captures the dynamic patient conditions as they progress through the waitlist journey.

*Random Survival Forest (RSF) Model:* We also employed a Random Survival Forest (RSF) (24) as a nonparametric alternative to Cox-based approaches. RSF is an ensemble learning method that extends the concept of random forests to right-censored survival data. It constructs a collection of survival trees, each grown on a bootstrap sample of the data, and aggregates predictions across the forest. RSF estimates conditional cumulative hazard function:

$$h\left(t \middle| x(s)\right) = \sum_{j:t_j \le t} \frac{d_j(x(s))}{Y_j(x(s))}$$

where $h\left(t \middle| x(t)\right)$ represents the cause-specific hazard at the time $t$ given the time-dependent covariates $x(s)$, $d_j(x(s))$ denotes the number of events at time $t_j$ for patients with covariates $x(s)$, $Y_j(x(s))$ denotes the number at risk at time $t_j$, and $\beta$ is the vector of regression coefficients. Likewise, this formulation allows RSF to track dynamic changes in patient condition over time by associating updated covariate values to future event risk.

*DeepHit Model:* We employed DeepHit model (25) to analyze waitlist outcomes, which processes longitudinal patient histories, sequences of covariate observations $x(s)$ at multiple past time points $s < t$ and predict the discrete-time probability of an event occurring at future time $t$. Mathematically, the survival function is expressed as:

$$S(t|N_s) = \sum_{r=t+1}^{T} p_r(N_s)$$

where $N_s = \{x(s_1), x(s_2), \ldots, x(s_k)\}$ denotes the covariate history up to time $s$, and $p_t(N_s) = P(T = t|N_s)$ denotes the predicted probability that the event occurs at time $t$. The model is trained using both likelihood and ranking losses, and it captures non-proportional and time-dependent survival risk patterns through a deep neural network architecture.

**Results**

In this study, we systematically reproduce models and variables from several prior works using the most recent UNOS dataset, benchmarking their performance across a range of evaluation metrics. Due to the nature of differences in data availability, a few variables used in previous studies had to be excluded. We also propose new models that incorporate a broader and longitudinal set of clinical variables. Our goal is to assess the predictive power of established methods and quantify the gain of our model by incorporating richer and time-varying data into waitlist mortality modeling.

After data preprocessing, we divided the 23,807 patients randomly into training-validation-test sets by an 80%-10%-10% ratio. ElasticNet was used for both regularization and feature selection. We performed a five-fold cross-validation grid search on the 90% training-validation split to identify the optimal hyperparameters (L1/L2 coefficients, learning rate, etc.) and evaluated model performance on the held-out test set. Predicted probabilities were recalibrated using logistic regression to fit the logit of the model's original risk estimates. A comprehensive set of metrics were used to evaluate model performance. C-index measures the model's ability to correctly rank patients by mortality risk over time. AUROC (AUC) at 1 year assesses the model's ability to distinguish between patients who will die or survive in 1 year across all classification thresholds. AUPRC (Area Under the Precision-Recall Curve) focuses on the model's ability to identify deaths among all predicted deaths, which is informative in imbalanced datasets. The most important metrics are AUROC, AUPRC, and C-index, as they capture ranking quality and discrimination performance under class imbalance. Brier Score at 1 year and Integrated Brier Score (IBS) in 1 year assess the accuracy and calibration of predicted survival probabilities. Recall and Precision at 1 year quantify the model's ability to correctly detect true deaths and minimize false positives. F1 Score balances these two. Accuracy is not so informative when classes are imbalanced. The baseline variables and longitudinally recorded variables in this study are shown in Table 1. Compared to previous models, we used a more comprehensive set of variables, especially including new longitudinal variables that are recorded continuously over time, covering richer information that was never been used before in the past.

Table 1. Summary statistics for the baseline variables (left) and longitudinally recorded variables (right) used in this study. Our model uses longitudinal variables: both baseline (at listing) and endline (last recorded) values are presented.

| Baseline Variables | Alshawabkeh | Jasseron | Hsich | Bakhtiyar | Ours |
|---|---|---|---|---|---|
| Age (years) | ✓ | | ✓ | ✓ | 52.9 ± 12.8 |
| Female gender | ✓ | | ✓ | ✓ | 6239 (26.2%) |
| Weight (kg) | | | | | 85.5 ± 18.8 |
| BMI | ✓ | | ✓ | ✓ | 28.1 ± 5.0 |
| Race | | | | | |
| Black | | | ✓ | | 6257 (26.3%) |
| White | | | ✓ | | 13895 (58.4%) |
| Hispanic | | | ✓ | | 2504 (10.5%) |
| Asian | | | ✓ | | 851 (3.6%) |
| Education | | | | | |
| Grade/High School | | | | | 9290 (39.0%) |
| College Study | | | | | 11261 (47.3%) |
| Graduate Degree | | | | | 2057 (8.6%) |
| Citizenship | | | | | |
| U.S. Citizen | | | | | 22681 (95.3%) |
| Non-U.S. Citizen | | | | | 969 (4.0%) |
| Type of dialysis | | | | | |
| Hemodialysis | | | | | 899 (3.8%) |
| Peritoneal Dialysis | | | | | 112 (0.5%) |
| Initial UNOS status | | | | | |
| OLD UNOS Status | ✓ | | | ✓ | 2668 (11.2%) |
| NEW UNOS Status | | | | | 20877 (87.7%) |
| Primary diagnosis | | | | | |
| Congenital heart disease | | | ✓ | | 1027 (4.3%) |
| Dilated cardiomyopathy | | | ✓ | ✓ | 7473 (31.4%) |
| Hypertrophic cardiomyopathy | | | ✓ | | 735 (3.1%) |
| Idiopathic cardiomyopathy | | | | ✓ | 8223 (34.5%) |
| Ischemic cardiomyopathy | | | ✓ | ✓ | 7047 (29.6%) |
| Restrictive cardiomyopathy | | | ✓ | | 1028 (4.3%) |
| Transplant graft failure/rejection | | | | | 515 (2.2%) |
| Valvular cardiomyopathy | | | ✓ | | 253 (1.1%) |
| Diabetes mellitus | ✓ | ✓ | ✓ | ✓ | 7266 (30.5%) |
| Cerebrovascular disease | | | ✓ | | 1798 (7.6%) |
| Previous cardiac surgery | ✓ | | | | 8804 (37.0%) |
| History of cigarette use | | | ✓ | | 9769 (41.0%) |
| Previous malignancy | | | ✓ | | 2126 (8.9%) |
| On inotropes | ✓ | | ✓ | ✓ | 7246 (30.4%) |
| On prostaglandins | | | | | 37 (0.2%) |
| Mechanical ventilation | ✓ | | ✓ | ✓ | 438 (1.8%) |
| ICD | ✓ | | ✓ | | 15825 (66.5%) |
| IABP | | ✓ | ✓ | ✓ | 2976 (12.5%) |
| ECMO | | ✓ | ✓ | ✓ | 884 (3.7%) |
| VAD (MCS) | ✓ | ✓ | | ✓ | 6717 (28.4%) |
| RVAD | | | | | 54 (0.2%) |
| LVAD | | | ✓ | | 6370 (26.8%) |
| TAH | | | | | 55 (0.2%) |
| RVAD+LVAD | | | | | 238 (1.0%) |
| Functional status (%) | | | | ✓ | 45.2 ± 23.0 |
| Number of prior transplants | | | | | 0.03 ± 0.18 |
| PVR (wood units) | ✓ | | | | 2.5 ± 2.0 |
| SPP (mmHg) | | | | | 14.1 ± 6.4 |
| Mean PAP | | | ✓ | | 27.1 ± 10.6 |
| Cardiac index | | | ✓ | | 2.2 ± 0.7 |
| TPG | | | | | 9.6 ± 6.5 |
| PVO2 | ✓ | | ✓ | | 60.3 ± 10.5 |
| PCWP | ✓ | | ✓ | | 17.6 ± 8.7 |
| Creatinine clearance (mL/min) | | | | | 85.5 ± 60.1 |
| Absolute creatinine (mg/dl) | | | | | 1.5 ± 1.3 |
| eGFR | | ✓ | ✓ | | 68.5 ± 28.7 |
| *Blood group* | | | | | |
| A | ✓ | | ✓ | | 8547 (35.9%) |
| B | ✓ | | ✓ | | 3390 (14.2%) |
| AB | ✓ | | ✓ | | 1000 (4.2%) |
| O | ✓ | | ✓ | | 10870 (45.7%) |

| Longitudinally Recorded Variable | Baseline (record at listing) | Endline (last record) |
|---|---|---|
| History of Stroke | 3151 (13.2%) | 3387 (14.2%) |
| History of Peripheral Thromboembolic Events | 1957 (8.2%) | 2118 (8.9%) |
| On Anti-Arrhythmics | 10509 (44.1%) | 10520 (44.2%) |
| On Continuous Invasive Mechanical Ventilation | 556 (2.3%) | 808 (3.4%) |
| On Dialysis | 1041 (4.4%) | 1189 (5.0%) |
| On a Diuretic | 18057 (75.8%) | 17102 (71.8%) |
| On Pulmonary Vasodilators | 2365 (9.9%) | 2555 (10.7%) |
| On Vasoactive Support | 9633 (40.5%) | 10381 (43.6%) |
| Support (Device/Inotrope) | | |
| Device Support | 6520 (27.4%) | 6967 (29.3%) |
| Inotrope Support | 7788 (32.7%) | 6334 (26.6%) |
| Inotrope Support and Device Support | 2550 (10.7%) | 3557 (14.9%) |
| On Oral Anticoagulant when INR was Obtained | 9954 (41.8%) | 9687 (40.7%) |
| BNP Test Type | | |
| BNP | 10976 (46.1%) | 10819 (45.4%) |
| NT Pro BNP | 9963 (41.8%) | 10120 (42.5%) |
| Number of Hospital Admissions in 12 Months | 1.616 ± 1.603 | 1.690 ± 1.679 |
| Total Number of Prior Sternotomies | 0.576 ± 0.819 | 0.616 ± 0.832 |
| Cardiac Index (in L/min/$m^2$) | 2.200 ± 0.707 | 2.236 ± 0.692 |
| Cardiac Output (in L/min) | 4.386 ± 1.335 | 4.462 ± 1.365 |
| Central Venous Pressure (in mmHg) | 9.276 ± 5.974 | 9.339 ± 6.006 |
| Diastolic Blood Pressure (in mmHg) | 69.73 ± 12.43 | 69.34 ± 12.58 |
| Mean Pulmonary Artery Pressure (in mmHg) | 27.39 ± 10.41 | 27.14 ± 10.46 |
| Pulmonary Artery Diastolic Pressure (in mmHg) | 19.40 ± 8.66 | 19.26 ± 8.70 |
| Pulmonary Artery Systolic Pressure (in mmHg) | 40.36 ± 14.65 | 39.86 ± 14.62 |
| PCWP (in mmHg) | 17.71 ± 8.74 | 17.62 ± 8.75 |
| Resting Heart Rate (in bpm) | 83.61 ± 19.09 | 84.00 ± 19.38 |
| Systolic Blood Pressure (in mmHg) | 105.6 ± 17.6 | 104.7 ± 17.8 |
| Mixed Venous Oxygen Saturation (SvO2) (in %) | 60.27 ± 10.35 | 60.29 ± 10.52 |
| Serum Albumin (g/dL) | 3.815 ± 0.631 | 3.775 ± 0.630 |
| AST (U/L) | 38.03 ± 91.43 | 39.20 ± 110.97 |
| Serum Bilirubin (mg/dL) | 1.004 ± 1.083 | 0.962 ± 1.131 |
| BNP (pg/mL) | 2682.8 ± 5299.7 | 2666.8 ± 5357.6 |
| BUN (mg/dL) | 25.39 ± 15.24 | 25.14 ± 15.23 |
| Serum Creatinine (mg/dL) | 1.490 ± 1.316 | 1.498 ± 1.374 |
| INR | 1.635 ± 0.788 | 1.668 ± 0.787 |
| Serum Sodium (mEq/L) | 136.7 ± 4.1 | 136.5 ± 4.2 |
| CPRA (%) | 11.40 ± 24.29 | 11.74 ± 24.65 |
| Sensitization - MFI Threshold (MFI) | 3519.4 ± 2693.0 | 3507.9 ± 2610.7 |

*Notes: Longitudinal variables are only used in "Ours" for the first time, not in other baseline models. Variables with >30% missingness are removed and not presented. Variable names follow UNOS definitions. For example, a patient may experience stroke on the waitlist, thus history of stroke can change over time.*

*Variable abbreviations are explained below.*

MCS: Mechanical Circulatory Support
PCWP: Pulmonary capillary wedge pressure
CPRA: Calculated panel reactive antibody
INR: International normalized ratio
MFI: Mean fluorescence intensity
BUN: Blood urea nitrogen
BNP: Brain Natriuretic Peptide
SPP: Systemic Perfusion Pressure
PAP: Pulmonary artery pressure
PVO2: Peak oxygen consumption
TPG: Transpulmonary gradient
TAH: Total artificial heart
PVR: Pulmonary Vascular Resistance
VAD: Ventricular assist device
LVAD: Left Ventricular assist device
RVAD: Right Ventricular assist device
IABP: Intra-Aortic Balloon Pump
ICD: Implantable cardiac defibrillator
eGFR: estimated Glomerular Filtration Rate
AST: Aspartate aminotransferase

Table 2. Performance comparison of all models on the new UNOS database. The best-performing model is highlighted in bold, while the second-best model is underlined. '1Y' represents one-year discrimination. Lower Brier Score and IBS indicate better performance, while higher values are better for all other metrics.

| Approach | # Features | Accuracy (1Y) | Precision (1Y) | Recall (1Y) | F1 Score (1Y) | Brier Score (1Y) | IBS (in 1 Year) | AUROC (1Y) | AUPRC (1Y) | C-Index |
|---|---|---|---|---|---|---|---|---|---|---|
| Alshawabkeh (2016) | 14 | 0.74±0.01 | 0.64±0.01 | 0.29±0.01 | 0.40±0.00 | 0.18±0.00 | 0.20±0.00 | 0.72±0.00 | 0.55±0.00 | 0.75±0.00 |
| Jasseron (2017) | 5 | 0.70±0.01 | 0.60±0.00 | 0.15±0.01 | 0.24±0.00 | 0.19±0.00 | 0.22±0.00 | 0.70±0.00 | 0.48±0.00 | 0.69±0.00 |
| Hsich (2018) | 21 | 0.75±0.01 | 0.66±0.00 | 0.32±0.01 | 0.43±0.00 | 0.17±0.00 | 0.20±0.00 | 0.75±0.00 | 0.58±0.00 | 0.79±0.00 |
| Bakhtiyar (2020) | 12 | 0.73±0.01 | 0.68±0.01 | 0.17±0.00 | 0.27±0.00 | 0.19±0.00 | 0.23±0.00 | 0.69±0.00 | 0.51±0.00 | 0.69±0.00 |
| Our Dynamic CPH | 77 | 0.87±0.01 | 0.76±0.01 | 0.42±0.00 | 0.54±0.00 | 0.11±0.00 | 0.17±0.00 | 0.87±0.00 | 0.66±0.00 | 0.87±0.00 |
| Our Static CPH | 43 | 0.78±0.01 | 0.72±0.01 | 0.40±0.00 | 0.52±0.00 | 0.16±0.00 | 0.20±0.00 | 0.76±0.00 | 0.63±0.00 | 0.79±0.00 |
| Our Dynamic RSF | 77 | **0.88±0.01** | 0.79±0.01 | 0.44±0.01 | 0.56±0.00 | 0.10±0.00 | 0.16±0.00 | 0.88±0.00 | 0.70±0.00 | 0.91±0.00 |
| Our Static RSF | 43 | 0.79±0.01 | 0.72±0.01 | 0.42±0.00 | 0.52±0.01 | 0.15±0.00 | 0.19±0.00 | 0.78±0.00 | 0.66±0.00 | 0.77±0.01 |
| Our Dynamic DeepHit | 77 | 0.87±0.01 | **0.80±0.01** | **0.45±0.01** | **0.57±0.01** | **0.10±0.00** | **0.16±0.00** | **0.89±0.00** | **0.70±0.00** | **0.94±0.01** |
| Our Static DeepHit | 43 | 0.80±0.01 | 0.73±0.01 | 0.42±0.01 | 0.52±0.00 | 0.15±0.00 | 0.19±0.00 | 0.77±0.00 | 0.65±0.00 | 0.80±0.00 |

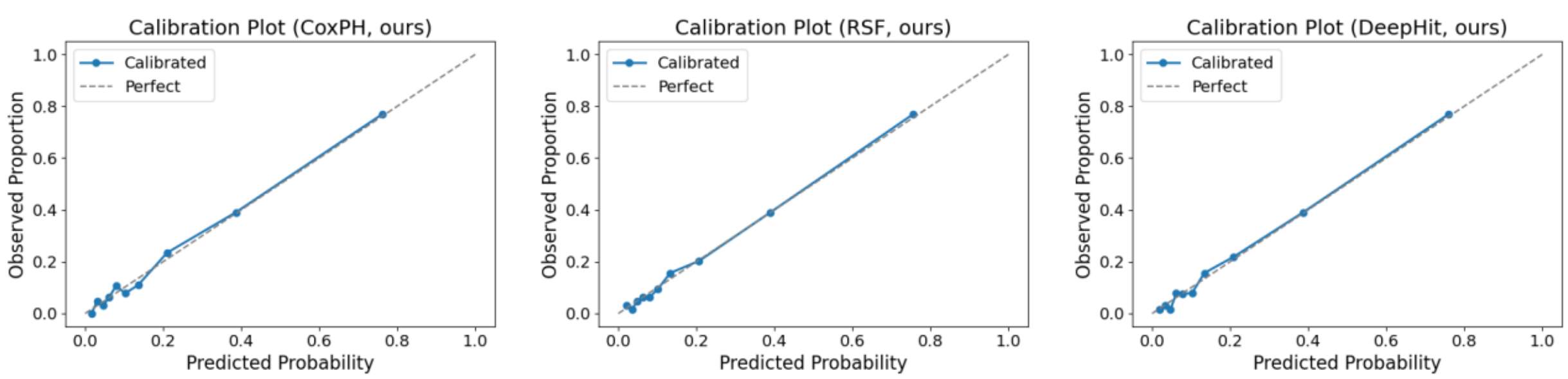

Figure 1. Calibration plots comparing our dynamic Cox model, Random Survival Forest, and DeepHit.

A comparison of the model results is shown in Table 2. We implement all previous baselines using the new UNOS dataset. Our "static" models (models trained using the baseline variables recorded at listing only) perform significantly better than previous baselines on almost all metrics. More interestingly, our "dynamic" models (models trained using the longitudinally recorded variables) perform much better than our "static" models, which demonstrates the value of the new longitudinal data that reflects the continuously evolving status of the patients. Among our three models, Cox Proportional Hazard (CPH) model performs the worst due to its linearity, while Random Survival Forest (RSF) model and DeepHit model are both able to capture nonlinear relationships. DeepHit seems to perform slightly better.

Across all models, recall is consistently lower than accuracy. As only 8% of patients died on the waitlist, this indicates that models are missing a substantial number of true positive cases. Class re-weighting leads to minimal impact on performance, likely because probability recalibration has already been used, which mitigated class imbalance effects. Overall, the variation in model performance under five-fold MCMC-based data imputation was negligible, suggesting a robust imputation process. The calibration plots for our dynamic models are shown in Figure 1, which evaluates how well the predicted survival probabilities align with the actual observed outcomes. Across all models, the predicted probabilities closely follow the diagonal line, indicating that models produce well-calibrated and reliable risk estimates.

Figure 2 compares the Cox baseline survival curve for a hypothetical patient with covariates set to zero against the Kaplan-Meier estimate for the overall cohort. The lower Cox curve indicates that the baseline patient has higher estimated mortality risk than the average patient. Among all the patients, 1,898 (8.0%) died waiting for a transplant. Based on the Kaplan-Meier analysis, the estimated survival probabilities at 1, 3, 5, and 10 years were 89%, 78%, 67%, and 50%, respectively. We also analyzed the effects of variables and risk factors. The statistic summary of the Cox proportional hazard model based on the new UNOS dataset is shown in Table 3 and Figure 3. The significant variables are mostly longitudinally recorded variables, with only a few baseline variables such as patient age in years and initial UNOS status. The effects of these significant variables align with existing medical evidence in the literature (11–13,21,22,26–29). For example, aging, renal dysfunction (creatinine, BUN and dialysis), inotropes, ventilation, ECMO, VAD, hospitalizations, anti-arrhythmics, pulmonary artery diastolic pressure, and higher INR are all associated with higher mortality, while serum albumin, diastolic blood pressure, and eGFR are associated with lower mortality. Some new variables lack clear evidence in existing literature, which suggests that our findings may be novel but require further validation. For example, AST, BNP test type, oral anticoagulants, SvO2, and number of previous transplants are rarely discussed. External validation is needed to confirm the significance and generalizability of these associations.

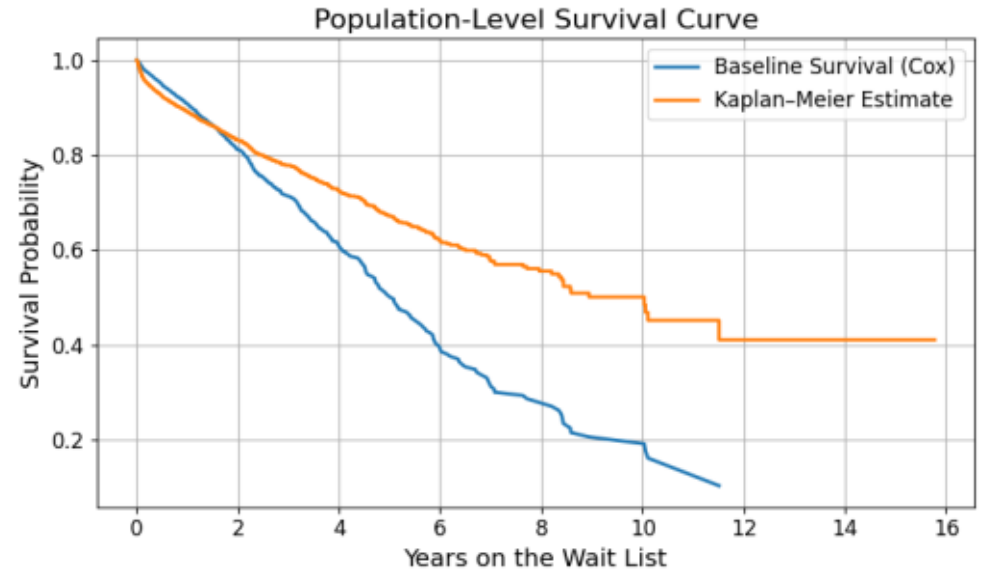

Figure 2. The baseline survival curve based on our Cox proportional hazard model and the Kaplan-Meier estimate.

Table 3. Summary of time-varying Cox proportional hazard model for heart transplant waitlist mortality. Significant variables with a p value less than 0.05 are listed. Variables with a positive impact on the mortality risk are marked by orange color, while variables with a negative impact are marked by blue color.

| Variable | Coef. | H.R. | Std. Err. | p value |
|---|---|---|---|---|
| Number of Hospital Admissions in 12 Months | 0.119 | 1.127 | 0.023 | 0 |
| On Anti-Arrhythmics: Yes | 0.161 | 1.175 | 0.046 | 0 |
| On Continuous Invasive Mechanical Ventilation: Yes | 0.815 | 2.258 | 0.112 | 0 |
| On Dialysis: Yes | 0.409 | 1.505 | 0.106 | 0 |
| Patient number of previous TXPs | -0.15 | 0.86 | 0.037 | 0 |
| Patient age in years | 0.314 | 1.369 | 0.029 | 0 |
| On Vasoactive Support: Yes | 0.472 | 1.603 | 0.065 | 0 |
| On Oral Anticoagulant when INR was Obtained: Yes | -0.323 | 0.724 | 0.058 | 0 |
| Serum Sodium (mEq/L) | -0.208 | 0.812 | 0.022 | 0 |
| Patient initial waitlist status: New Status 1 | 0.791 | 2.206 | 0.17 | 0 |
| BUN (mg/dL) | 0.108 | 1.114 | 0.019 | 0 |
| AST (U/L) | 0.026 | 1.027 | 0.007 | 0 |
| Serum Albumin (g/dL) | -0.343 | 0.71 | 0.022 | 0 |
| Serum Bilirubin (mg/dL) | 0.063 | 1.065 | 0.009 | 0 |
| Patient initial waitlist status: New Status 2 | 0.418 | 1.519 | 0.091 | 0 |
| Patient initial waitlist status: Old Status 2 | -0.614 | 0.541 | 0.107 | 0 |
| Central Venous Pressure (in mmHg) | 0.156 | 1.169 | 0.025 | 0 |
| Resting Heart Rate (in bpm) | 0.104 | 1.109 | 0.024 | 0 |
| Mixed Venous Oxygen Saturation (SvO2) (in %) | -0.087 | 0.916 | 0.026 | 0.001 |
| Patient functional status at WL | -0.093 | 0.911 | 0.027 | 0.001 |
| BNP (pg/mL) | 0.056 | 1.058 | 0.018 | 0.002 |
| INR | 0.072 | 1.074 | 0.023 | 0.002 |
| Patient initial waitlist status: Old Status 1A | -0.485 | 0.616 | 0.164 | 0.003 |
| On Pulmonary Vasodilators: Yes | 0.189 | 1.208 | 0.066 | 0.004 |
| Patient On ECMO at REG: Yes | 0.462 | 1.587 | 0.163 | 0.005 |
| Diastolic Blood Pressure (in mmHg) | -0.077 | 0.926 | 0.028 | 0.006 |
| Patient initial waitlist status: Old Status 1B | -0.259 | 0.772 | 0.095 | 0.007 |
| Patient eGFR | -0.076 | 0.927 | 0.031 | 0.013 |
| Patient initial waitlist status: New Status 3 | 0.248 | 1.281 | 0.1 | 0.014 |
| Patient initial waitlist status: New Status 6 | -0.191 | 0.826 | 0.085 | 0.026 |
| On a Diuretic: Yes | -0.117 | 0.89 | 0.054 | 0.032 |
| Pulmonary Artery Diastolic Pressure (in mmHg) | 0.08 | 1.083 | 0.039 | 0.041 |
| History of Stroke: Yes | 0.126 | 1.134 | 0.063 | 0.047 |

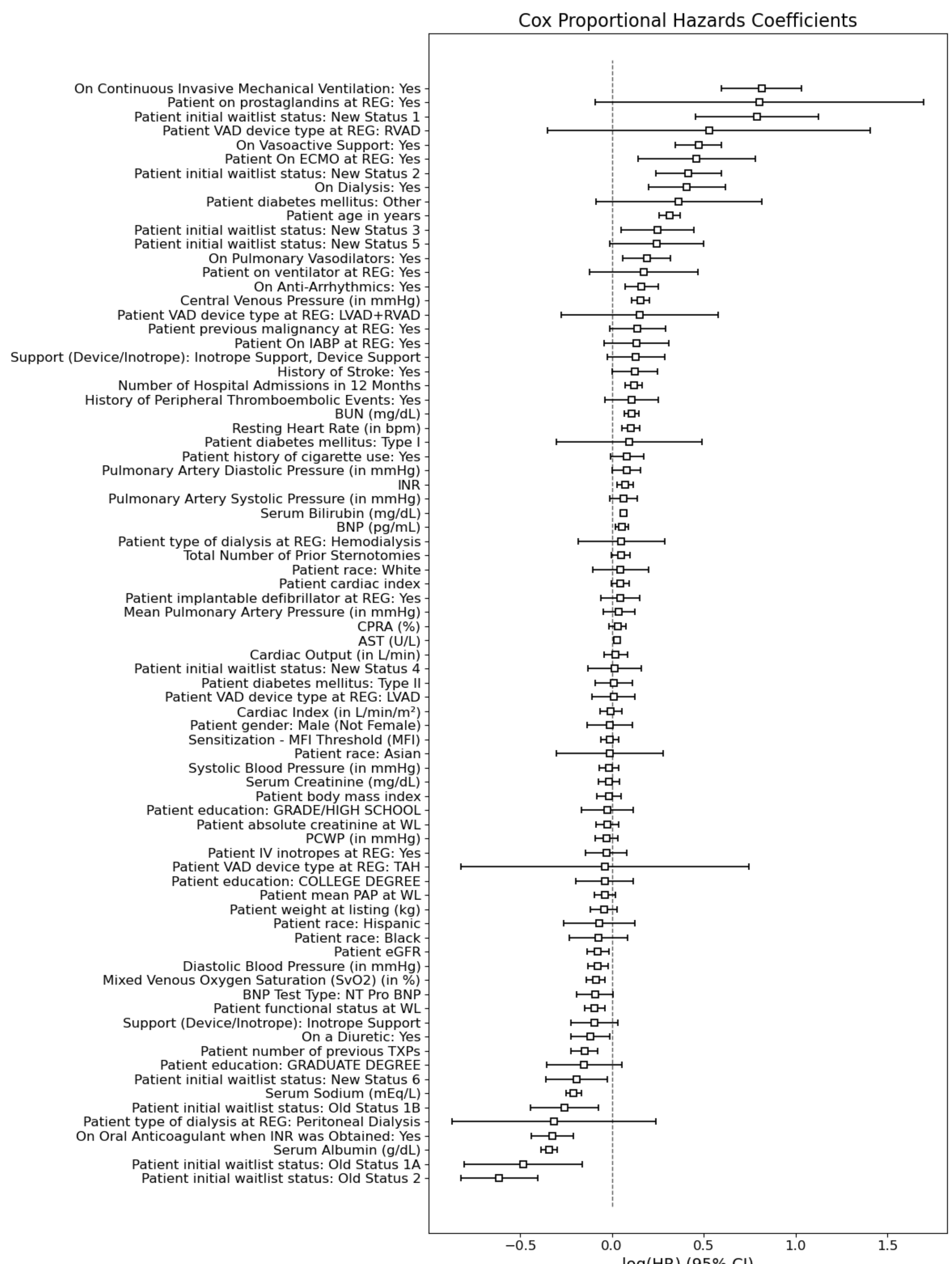

Figure 3. The log hazard ratio of all the variables in our Cox proportional hazard model, which directly reflects the effect (positive versus negative) under uncertainty of each covariate.

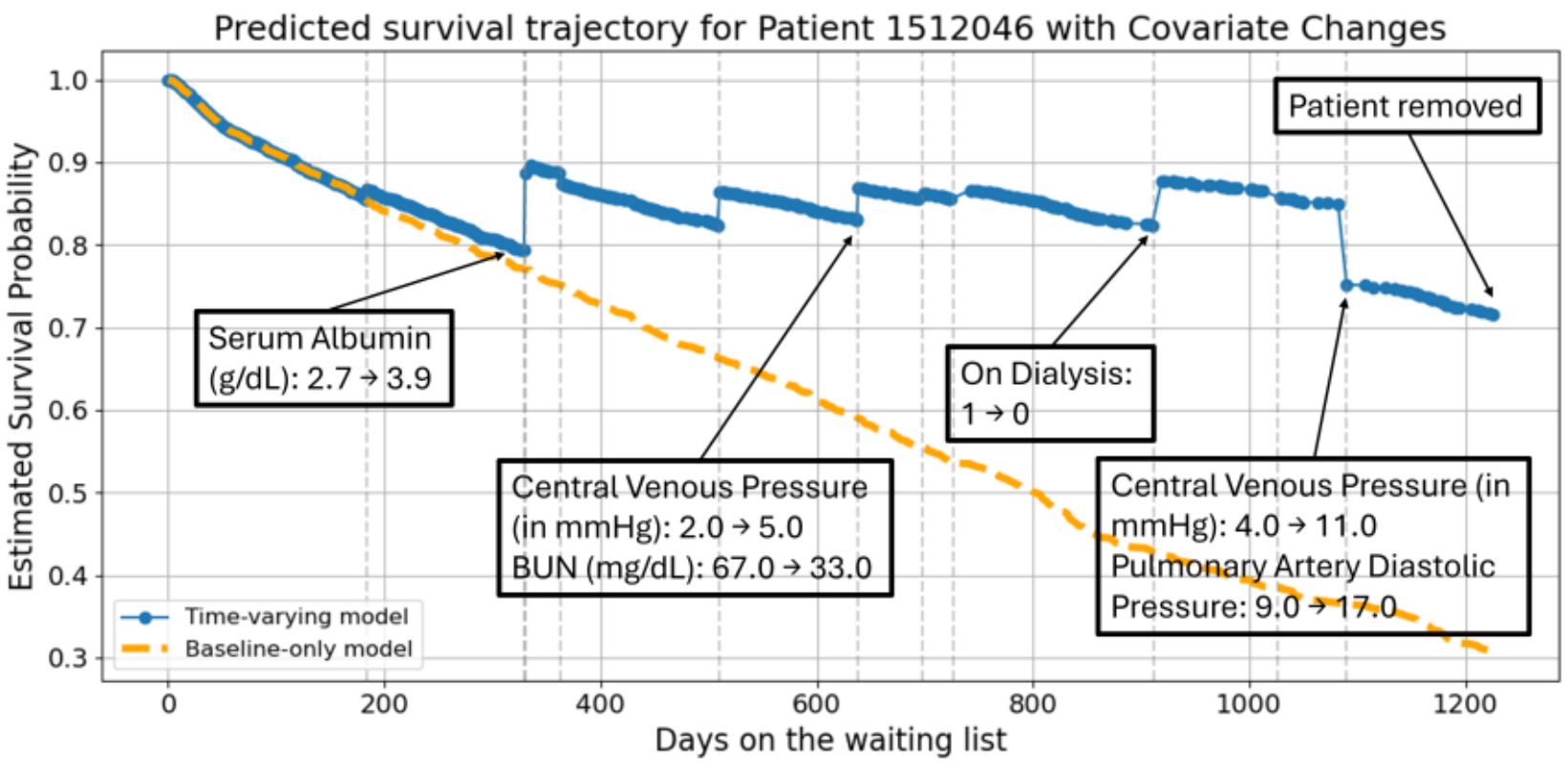

Figure 4. Individual risk profile for patient ID 1512046 who was removed from the wait list due to improved condition.

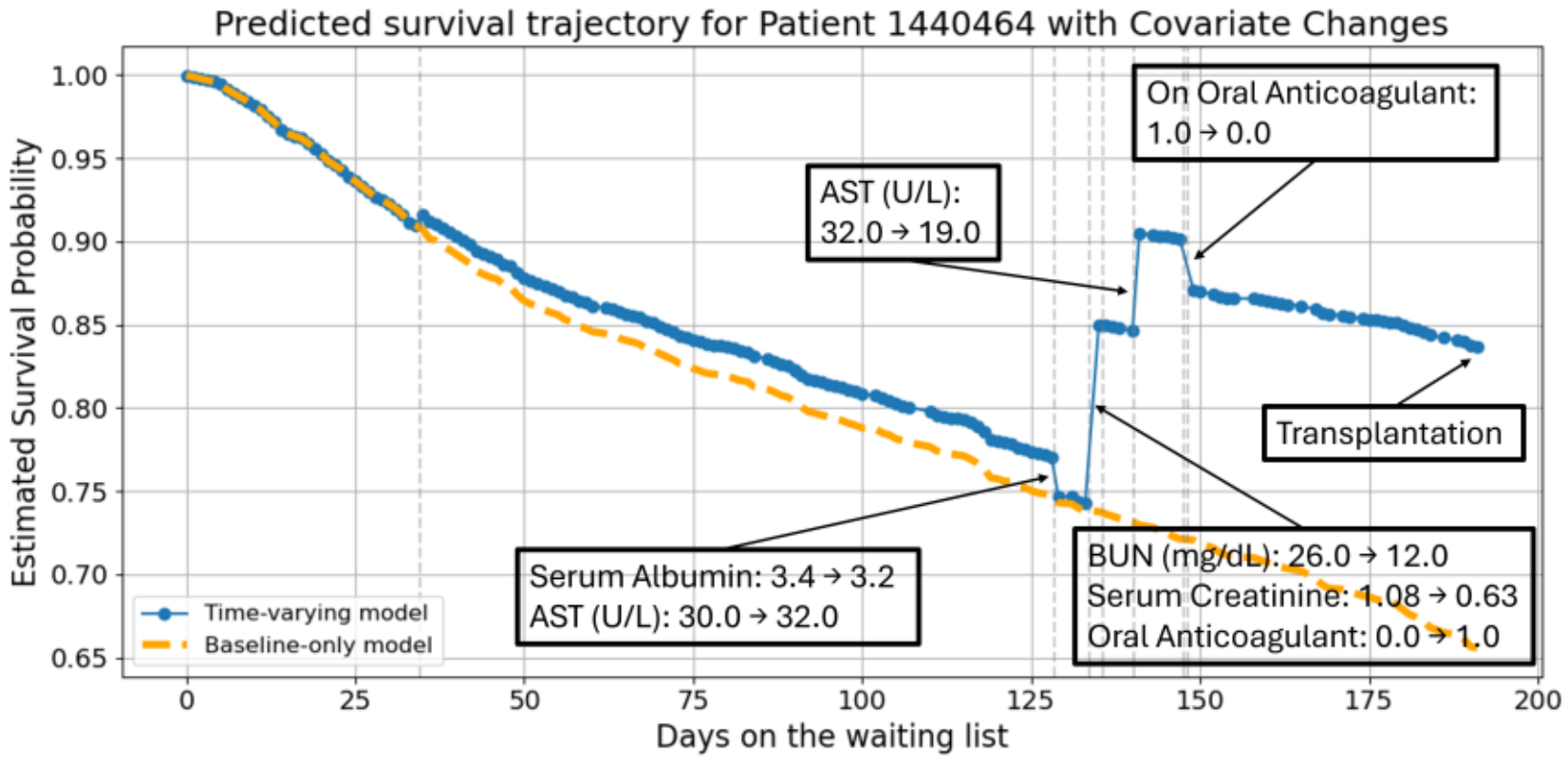

Figure 5. Individual risk profile for patient ID 1440464 who got a transplant at the end.

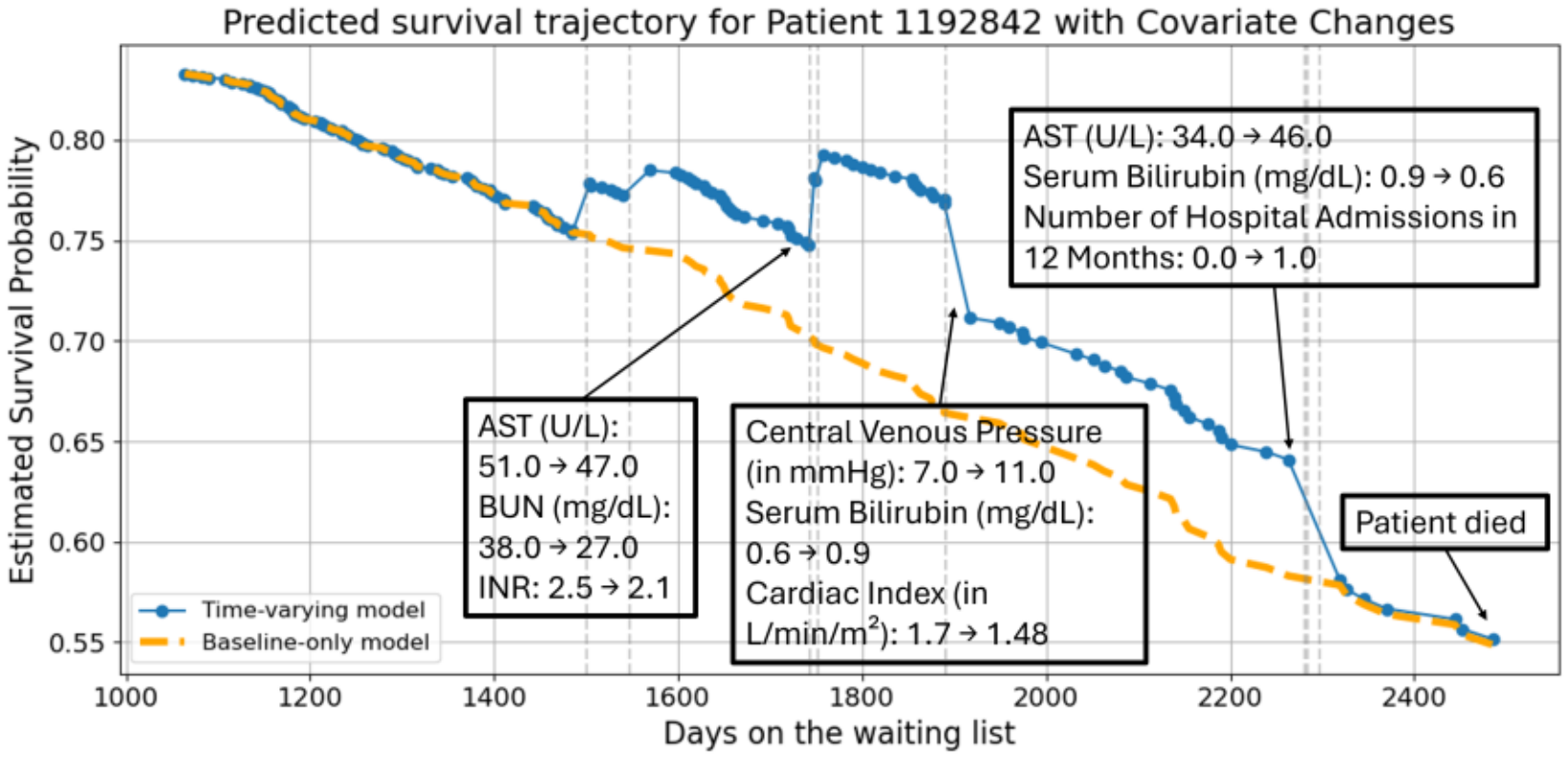

Figure 6. Individual risk profile for patient ID 1192842 who died on the waiting list at the end.

We illustrate the value of longitudinally recorded variables and time-varying modeling through three individual patient trajectories leading to removal, transplant, and death. In Figure 4, the patient exhibited a decline in BUN levels and was removed from dialysis, resulting in a marked rise in predicted survival probability and eventual removal from the waitlist without the need for a transplant any longer. In Figure 5, the patient was not on any device support, and the survival probability increased and stabilized at around 135 days. Then, the patient received an organ successfully. In Figure 6, the patient had several changes in critical risk factors such as AST, cardiac index, and serum Bilirubin. The patient had multiple donor offers but all of them were rejected. Later, the patient still did not match a donor offer and died. These examples highlight the benefit of incorporating longitudinal covariates, which enable real-time updates to risk predictions based on evolving clinical status, capturing critical dynamics that static baseline-only models miss.

## Discussion

Our results highlight the value of modeling time-varying patient data for predicting heart transplant waitlist mortality. Dynamic models outperformed static ones, indicating that evolving clinical status holds crucial prognostic signals. This aligns with a broader shift toward personalized, time-sensitive risk prediction in survival modeling. Our models identified established risk factors, e.g., elevated creatinine, BUN, low sodium, dialysis, and mechanical support, as well as less commonly reported markers in the literature, such as AST, SvO2, and oral anticoagulant use. These may reflect underrecognized markers of systemic health or patient management strategies and warrant further investigation. Notably, the number of prior transplants appeared to be associated with improved survival in our model, in contrast to earlier findings; this may reflect selection bias in patients.

Our findings suggest that UNOS could benefit from integrating dynamic models into clinical workflows to provide real-time updates to patient risk, improving urgency assessments and potentially reducing waitlist mortality. Adoption would require cross-center validation and careful attention to interpretability and algorithmic biases.

Several limitations should be noted. Our dataset, although large and rich, is observational and collected primarily for administrative and allocation purposes rather than prospective modeling. Some clinically relevant features were not captured, and missing data required imputation. Moreover, while our models showed strong internal validity, they have not yet been externally validated on data from other regions or healthcare systems.

Future work should explore external validation, integration of unstructured data such as clinical notes or imaging, and extension to related tasks such as time-to-transplant and time-to-donor-offer prediction. Ultimately, combining high-resolution, time-varying data with interpretable, deployable modeling techniques could support more personalized and equitable heart transplant decision-making.

## Conclusion

This study establishes the first benchmark for heart transplantation waitlist mortality prediction, spanning traditional linear models, machine learning methods, and deep learning-based survival models. We evaluate models using not only baseline variables recorded only at the time of listing, but also longitudinal variables that evolve throughout the waitlist period. Our results show that incorporating time-varying data leads to substantial gains in model predictive performance, with the best dynamic model achieving a C-Index of 0.94 and a 1-year AUROC of 0.89. Analysis on the effects and significance of variables was conducted, and we identified that the new longitudinally recorded clinical variables are generally more significant than clinical variables recorded only at listing. The effects of most variables align with the existing literature, but further verification is needed. These findings highlight the value of dynamic modeling and lay the groundwork for future clinical decision support systems that continuously update patient risk estimation and enable a timely, data-driven heart transplant decision-making opportunity. Our benchmark provides a standardized reference for future work and emphasizes the critical role of longitudinal data.

## Acknowledgement

This research is supported by the Translational Fellowship in Digital Health from the Center for Machine Learning and Health at CMU and an NIH-NHLBI grant 1R01HL162882-01A1 to AK and RP. We also acknowledge valuable feedback from physicians at MUSC and data support from UNOS.